\RequirePackage{lineno} 
\documentclass[prl, twocolumn, superscriptaddress, showpacs]{revtex4}
\usepackage{epsfig}
\usepackage{color}
\usepackage{amsmath}

\usepackage{graphicx} 
\usepackage{dcolumn}  
\usepackage{bm}       
\usepackage{amsfonts, amssymb}
\newcommand{\gevsq}{GeV$^2$}

\newcommand{\qsq}{$Q^2$}
\newcommand{\cma}{$\theta_{\mathrm{cm}}^{\mathrm{p}}$}

\newcommand{\invs}{$s$}
\newcommand{\invt}{$t$}

\newcommand{\gep}{$G_{\mathrm{E}}^{\mathrm{p}}$}
\newcommand{\gmp}{$G_{\mathrm{M}}^{\mathrm{p}}$}
\newcommand{\KLL}{$K_{_{\mathrm{LL}}}$}
\newcommand{\KLS}{$K_{_{\mathrm{LS}}}$}
\newcommand{\ada}{$\phi_1(x_1,x_2,x_3) = 120 x_1x_2x_3$}
\newcommand{\ein}{$E_{\mathrm{inc}}$}
\newcommand{\PR}{{\em Phys. Rev. }}

\newcommand{\PRL}{{\em Phys. Rev. Lett. }}

\newcommand{\NP}{{\em Nucl. Phys. }}
\newcommand{\NIM}{{\em Nucl. Instrum. Methods }Phys. Res.,}
\newcommand{\JHEP}{{\em J. High Energy Phys. }}
\newcommand{\etal}{{\em et al.}}

\begin{document}

\title{Polarization Transfer in Wide-Angle Compton Scattering and Single-Pion Photoproduction from the Proton}
\author{C.~Fanelli} \affiliation{\mbox{Dipartimento di Fisica, Universit\`{a}  La Sapienza, Rome, Italy
and INFN, Sezione di Roma, 00185, Rome, Italy}}
\affiliation{\mbox{INFN, Sezione di Roma, gruppo Sanit\`{a} 
and Istituto Superiore di Sanit\`{a} , 00161 Rome, Italy}} 
\author{E.~Cisbani} \affiliation{\mbox{INFN, Sezione di Roma, gruppo Sanit\`{a} 
and Istituto Superiore di Sanit\`{a} , 00161 Rome, Italy}} 
\author{D.~J.~Hamilton} 
\affiliation{\mbox{University of Glasgow, Glasgow G12 8QQ, Scotland, United Kingdom}}
\author{G.~Salm\'{e}} \affiliation{\mbox{Dipartimento di Fisica, Universit\`{a} La Sapienza, Rome, Italy
and INFN, Sezione di Roma, 00185, Rome, Italy}}
\author{B.~Wojtsekhowski} \thanks{Corresponding author: bogdanw@jlab.org}
\affiliation{\mbox{Thomas Jefferson National Accelerator Facility, Newport News, Virginia 23606, USA}}
\author{A.~Ahmidouch}
\affiliation{North Carolina A\&T State University, Greensboro, North Carolina 27411, USA}
\author{J.~R.~M.~Annand}
\affiliation{\mbox{University of Glasgow, Glasgow G12 8QQ, Scotland, United Kingdom}}
\author{H.~Baghdasaryan}
\affiliation{University of Virginia, Charlottesville, Virginia 22904, USA}
\author{J.~Beaufait}
\affiliation{\mbox{Thomas Jefferson National Accelerator Facility, Newport News, Virginia 23606, USA}}
\author{P.~Bosted}
\affiliation{\mbox{Thomas Jefferson National Accelerator Facility, Newport News, Virginia 23606, USA}}
\author{E.~J.~Brash}
\affiliation{Christopher Newport University, Newport News, Virginia 23606, USA}
\affiliation{\mbox{Thomas Jefferson National Accelerator Facility, Newport News, Virginia 23606, USA}}
\author{C.~Butuceanu}
\affiliation{University of Regina, Regina, Saskatchewan S4S OA2, Canada}
\author{P.~Carter}
\affiliation{Christopher Newport University, Newport News, Virginia 23606, USA}
\author{E.~Christy}
\affiliation{Hampton University, Hampton, Virginia 23668, USA}
\author{E.~Chudakov}
\affiliation{\mbox{Thomas Jefferson National Accelerator Facility, Newport News, Virginia 23606, USA}}
\author{S.~Danagoulian}
\affiliation{North Carolina A\&T State University, Greensboro, North Carolina 27411, USA}
\author{D.~Day}
\affiliation{University of Virginia, Charlottesville, Virginia 22904, USA}
\author{P.~Degtyarenko}
\affiliation{\mbox{Thomas Jefferson National Accelerator Facility, Newport News, Virginia 23606, USA}}
\author{R.~Ent}
\affiliation{\mbox{Thomas Jefferson National Accelerator Facility, Newport News, Virginia 23606, USA}}
\author{H.~Fenker}
\affiliation{\mbox{Thomas Jefferson National Accelerator Facility, Newport News, Virginia 23606, USA}}
\author{M.~Fowler}
\affiliation{\mbox{Thomas Jefferson National Accelerator Facility, Newport News, Virginia 23606, USA}}
\author{E.~Frlez}
\affiliation{University of Virginia, Charlottesville, Virginia 22904, USA}
\author{D.~Gaskell}
\affiliation{\mbox{Thomas Jefferson National Accelerator Facility, Newport News, Virginia 23606, USA}}
\author{R.~Gilman}
\affiliation{\mbox{Thomas Jefferson National Accelerator Facility, Newport News, Virginia 23606, USA}}
\affiliation{Rutgers, The State University of New Jersey,  Piscataway, New Jersey 08855, USA}
\author{T.~Horn}
\affiliation{\mbox{Thomas Jefferson National Accelerator Facility, Newport News, Virginia 23606, USA}}
\author{G.~M.~Huber}
\affiliation{University of Regina, Regina, Saskatchewan S4S OA2, Canada}
\author{C.~W.~de~Jager}
\affiliation{\mbox{Thomas Jefferson National Accelerator Facility, Newport News, Virginia 23606, USA}}
\affiliation{University of Virginia, Charlottesville, Virginia 22904, USA}
\author{E.~Jensen}
\affiliation{Christopher Newport University, Newport News, Virginia 23606, USA}
\author{M.~K.~Jones}
\affiliation{\mbox{Thomas Jefferson National Accelerator Facility, Newport News, Virginia 23606, USA}}
\author{A.~Kelleher}
\affiliation{College of William and Mary, Williamsburg, Virginia 23187, USA}
\author{C.~Keppel}
\affiliation{Hampton University, Hampton, Virginia 23668, USA}
\author{M.~Khandaker}
\affiliation{Norfolk State University, Norfolk, Virginia 23504, USA}
\author{M.~Kohl}
\affiliation{Hampton University, Hampton, Virginia 23668, USA}
\author{G.~Kumbartzki}
\affiliation{Rutgers, The State University of New Jersey,  Piscataway, New Jersey 08855, USA}
\author{S.~Lassiter}
\affiliation{\mbox{Thomas Jefferson National Accelerator Facility, Newport News, Virginia 23606, USA}}
\author{Y.~Li}
\affiliation{Hampton University, Hampton, Virginia 23668, USA}
\author{R.~Lindgren}
\affiliation{University of Virginia, Charlottesville, Virginia 22904, USA}
\author{H.~Lovelace}
\affiliation{Norfolk State University, Norfolk, Virginia 23504, USA}
\author{W.~Luo}
\affiliation{Lanzhou University, Lanzhou 730000, Gansu, People's Republic of China}
\author{D.~Mack}
\affiliation{\mbox{Thomas Jefferson National Accelerator Facility, Newport News, Virginia 23606, USA}}
\author{V.~Mamyan}
\affiliation{University of Virginia, Charlottesville, Virginia 22904, USA}
\author{D.~J.~Margaziotis}
\affiliation{California State University Los Angeles, Los Angeles, California 90032, USA}
\author{P.~Markowitz}
\affiliation{Florida International University, Miami, Florida 33199, USA}
\author{J.~Maxwell}
\affiliation{University of Virginia, Charlottesville, Virginia 22904, USA}
\author{G.~Mbianda}
\affiliation{University of Witwatersrand, Johannesburg, South Africa}
\author{D.~Meekins}
\affiliation{\mbox{Thomas Jefferson National Accelerator Facility, Newport News, Virginia 23606, USA}}
\author{M.~Meziane}
\affiliation{College of William and Mary, Williamsburg, Virginia 23187, USA}
\author{J.~Miller}
\affiliation{University of Maryland, College Park, Maryland 20742, USA}
\author{A.~Mkrtchyan}
\affiliation{Yerevan Physics Institute, Yerevan 375036, Armenia}
\author{H.~Mkrtchyan}
\affiliation{Yerevan Physics Institute, Yerevan 375036, Armenia}
\author{J.~Mulholland}
\affiliation{University of Virginia, Charlottesville, Virginia 22904, USA}
\author{V.~Nelyubin}
\affiliation{University of Virginia, Charlottesville, Virginia 22904, USA}
\author{L.~Pentchev}
\affiliation{College of William and Mary, Williamsburg, Virginia 23187, USA}
\author{C.~F.~Perdrisat}
\affiliation{College of William and Mary, Williamsburg, Virginia 23187, USA}
\author{E.~Piasetzky}
\affiliation{University of Tel Aviv, Tel Aviv, Israel}
\author{Y.~Prok}
\affiliation{Christopher Newport University, Newport News, Virginia 23606, USA}
\author{A.~J.~R.~Puckett} \affiliation{Massachusetts Institute of Technology, Cambridge, Massachusetts 02139, USA}
\author{V.~Punjabi}
\affiliation{Norfolk State University, Norfolk, Virginia 23504, USA}
\author{M.~Shabestari}
\affiliation{University of Virginia, Charlottesville, Virginia 22904, USA}
\author{A.~Shahinyan}
\affiliation{Yerevan Physics Institute, Yerevan 375036, Armenia}
\author{K.~Slifer}
\affiliation{University of Virginia, Charlottesville, Virginia 22904, USA}
\author{G.~Smith}
\affiliation{\mbox{Thomas Jefferson National Accelerator Facility, Newport News, Virginia 23606, USA}}
\author{P.~Solvignon}
\affiliation{Argonne National Laboratory, Argonne, Illinois 60439, USA}
\author{R.~Subedi}
\affiliation{University of Virginia, Charlottesville, Virginia 22904, USA}
\author{F. R.~Wesselmann}
\affiliation{Norfolk State University, Norfolk, Virginia 23504, USA}
\author{S.~Wood}
\affiliation{\mbox{Thomas Jefferson National Accelerator Facility, Newport News, Virginia 23606, USA}}
\author{Z.~Ye}
\affiliation{Hampton University, Hampton, Virginia 23668, USA}
\author{X.~Zheng}
\affiliation{University of Virginia, Charlottesville, Virginia 22904, USA}

\date{\today}

\begin{abstract}               

Wide-angle exclusive Compton scattering and single-pion
photoproduction from the proton have been investigated via
measurement of the polarization transfer from a circularly polarized
photon beam to the recoil proton.  The wide-angle Compton scattering
polarization transfer was analyzed at an incident photon energy of
3.7~GeV at a proton scattering angle of \cma$= 70^\circ$.  The
longitudinal transfer \KLL, measured to be $0.645 \pm 0.059 \pm
0.048$, where the first error is statistical and the second is
systematic, has the same sign as predicted for the reaction mechanism
in which the photon interacts with a single quark carrying the spin of
the proton.  However, the observed value is $\sim$3~times larger than
predicted by the generalized-parton-distribution-based calculations,
which indicates a significant unknown contribution to the scattering
amplitude.
  
\end{abstract}

\preprint{APS/123-QED}

\pacs{13.60.Fz, 14.20.Dh}

\maketitle


Understanding the structure of hadrons in terms of QCD is one of the
fundamental goals of modern nuclear physics.  The formalism of
generalized parton distributions (GPDs), developed about 20 years ago,
for the first time linked hadron structure information accessible
through inclusive reactions such as deep inelastic scattering to
information from exclusive reactions.  These GPDs, while not directly
measurable in experiments, provide a unified description of key
electromagnetic reactions on the nucleon~\cite{GPDintro}.  Whereas
deep inelastic scattering allows investigation of the longitudinal
structure of the nucleon, exclusive reactions such as elastic electron
and photon scattering access its transverse structure. Taken together
they allow determination of a complete image of the nucleon and its
complex substructure~\cite{GPD_hadron}.

Wide-angle Compton scattering (WACS) from the nucleon with large values of $s$, $-t$,
and $-u$ compared with $\Lambda_{_{\mathrm{QCD}}}^2$ is a hard
exclusive process that provides access to information about nucleon
structure that is complementary to high \qsq~elastic form factors and
deeply virtual Compton scattering. The common feature of these
reactions is a large energy scale, leading to factorization of the
scattering amplitude into a hard perturbative kernel and a factor
described by soft nonperturbative wave functions.

Various theoretical approaches have been applied to WACS in the
hard-scattering regime, and these can be distinguished by the number
of active quarks participating in the hard subprocess, or
equivalently, by the mechanism for sharing the transferred momentum
among the constituents.  Two extreme pictures have been proposed.  In
the perturbative QCD (pQCD) approach, three active quarks share the
transferred momentum by the exchange of two hard
gluons~\cite{fa90,kr91}.
In the handbag approach, which has in recent years become a staple in
the interpretation of data from hard exclusive reactions, only one
quark, whose wave function has sufficient high-momentum components for
the quark to absorb and reemit the photon~\cite{ra98,di99,hu02}, is
assumed to be active.  In any given kinematic regime both mechanisms
will contribute, in principle, to the cross section.  It is generally
believed that at sufficiently high energies the pQCD mechanism
dominates.  However, in the currently accessible experimental domain
of $s$ and $t$, the nature of the reaction mechanism is not fully
understood.

Three other theoretical advances based on leading-quark dominance in
WACS have been proposed in recent years. The constituent quark model
with a handbag diagram has proven successful in describing the WACS
process~\cite{mi04}, as have calculations performed in an extended
Regge model~\cite{ca03}.  More recently, the soft-collinear effective
theory (SCET) was developed for elastic electron-proton scattering at
high-momentum transfer~\cite{ki13}.  The QCD factorization approach
formulated in the framework of SCET allows for the development of a
description of the soft-spectator scattering contribution to the
overall amplitude.  The two-photon exchange contributions to elastic
electron-proton scattering were shown to factorize by the introduction
of a single, universal SCET form-factor which defines the dominant
soft-spectator amplitudes.  As it is argued in Ref.~\cite{ki13}, the
same form factor also arises naturally in WACS, and the most promising
route for understanding this soft spectator contribution in hard
exclusive reactions at JLab energies is the study of WACS.

One of the main predictions of the pQCD mechanism for WACS is the
constituent scaling rule~\cite{br73}, whereby $d\sigma/dt$ scales as
$s^{-6}$ at fixed \cma.  The pioneering experiment at
Cornell~\cite{sh79} was approximately consistent with constituent
scaling, albeit with modest statistical precision.  However, the
high-precision data from JLab gave a scaling power of $s^{-7.5 \pm
  0.2}$~\cite{da07}.  The calculations from both the GPD-based handbag
approach and the SCET framework have reproduced the JLab cross section
data very well.  Crucially, the extracted values of the SCET form
factor do not show any significant dependence on the value of $s$ as
required by factorization.

The longitudinal and sideways polarization transfer observables,
\KLL~and \KLS, respectively, are defined by:
\begin{eqnarray}
K_{\mathrm{_{LL}}} \, &\equiv& \,
\frac{d\sigma(+,\rightarrow)-d\sigma(-,\rightarrow)}
{d\sigma(+,\rightarrow)+d\sigma(-,\rightarrow)},
 \nonumber
\\
K_{\mathrm{_{LS}}} \, &\equiv& \,
\frac{d\sigma(+,\uparrow)-d\sigma(-,\uparrow)}
{d\sigma(+,\uparrow)+d\sigma(-,\uparrow)},
 \nonumber
\end{eqnarray}
where the first sign refers to the incident photon helicity and the
arrow to the recoil proton longitudinal ($\rightarrow$) or sideways
($\uparrow$) polarization.
The polarization transfer observables were previously measured at JLab
for Compton scattering at $s=6.9$ and $t=-4.1~$\gevsq~in experiment
E99-114~\cite{ha05}, whose concept is mainly repeated here at
different kinematics.  It was found that the longitudinal component of
the polarization transfer at the E99-114 kinematic point is large and
positive, in agreement with the handbag GPD and SCET predictions in
spite of a relatively low value of $u=-1.0~$\gevsq~and in unambiguous
disagreement with the pQCD predictions.
 
\begin{figure}[!h]
\vskip -0.05 in
\centering
\includegraphics[trim = 30mm 10mm 25mm 10mm, width = 0.37 \textwidth]{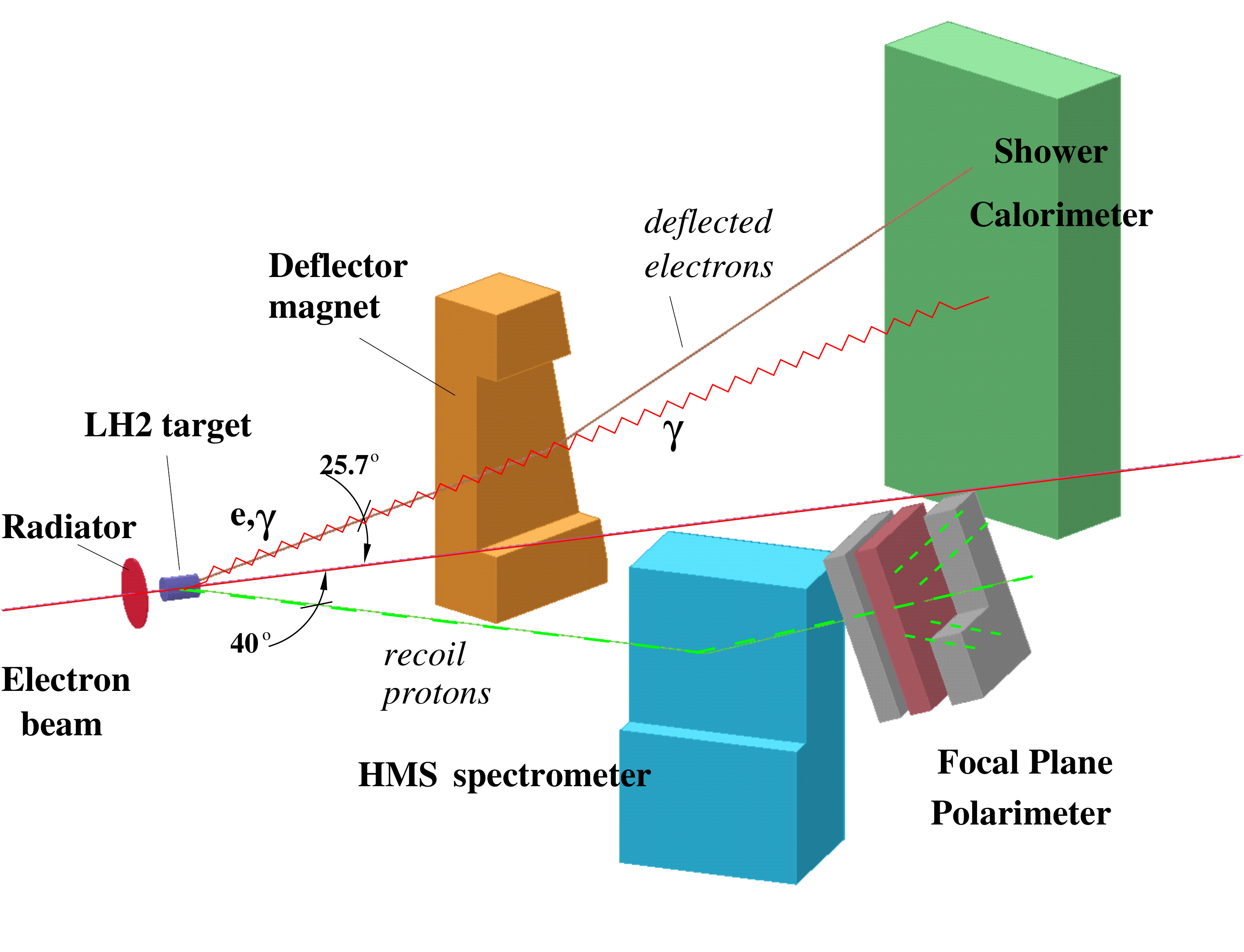}
\caption{Schematic layout of the E07-002 experiment.}
\label{fig:scheme}
\vskip -0.09 in
\end{figure}

The measurement reported in this Letter (JLab experiment E07-002) was
carried out in Hall C at Jefferson Lab, with the purpose of providing
values of \KLL~and \KLS~when all the Mandelstam variables are larger
than $\Lambda_{_{QCD}}^2$.
The layout of the experiment is shown schematically in
Fig.~\ref{fig:scheme}.  A longitudinally polarized, 100\% duty-factor
electron beam with current up to 40~$\mu$A and energy of 4.11~GeV was
incident on a copper radiator of 1.3~mm thickness placed on the beam
line.  The mixed beam of electrons and bremsstrahlung photons was
incident on a 15-cm liquid H$_2$ target, located just downstream from
the radiator, with a photon flux of up to $10^{13}$ equivalent
quanta/s.
For incident photons at an average weighted energy of 3.7~GeV, the
scattered photons were detected at a scattering angle of 25.7$^\circ$
in the BigCal calorimeter, which is composed of 1744 lead-glass bars
subtending a solid angle of 34~msr with an angular resolution of
1.8~mrad and relative energy resolution of 12\%.  The associated
recoil proton was detected in the Hall C High Momentum Spectrometer
(HMS) at the corresponding central angle of 40$^\circ$ and central
momentum of 1.85~GeV.  The proton was detected within a solid angle of
5~msr and momentum acceptance of $\pm$~9\%.  The trigger was formed
from a coincidence between a signal from scintillator counters in the
HMS and a signal above a 500~MeV threshold in the calorimeter.  A
magnet between the target and the calorimeter, as shown in
Fig.~\ref{fig:scheme}, with $\int \vec{B}\times \vec{dl} = 1.2$~Tm
deflected the elastically scattered electrons vertically by $\sim$
-50~cm relative to undeflected WACS photons.  Events with a radiative
photon kinematically indistinguishable from WACS constitute an
irreducible background.

Data have been collected with the radiator present and removed, and
with different field settings of the deflection magnet.  About 7.4~C
of beam charge was accumulated for WACS production runs.  The electron
beam longitudinal polarization was found to be $75.0 \pm 1.1$\% using
a M\o{}ller polarimeter.  During data taking, the beam polarization
was flipped at a 30~Hz rate.  The bremsstrahlung photon has 99\% of
the initial electron polarization over the energy range of current
analysis.

\begin{figure}[!htb]
\vskip -0.00 in
\includegraphics[trim = 100mm 20mm 0mm 25mm, width = 0.52 \textwidth]{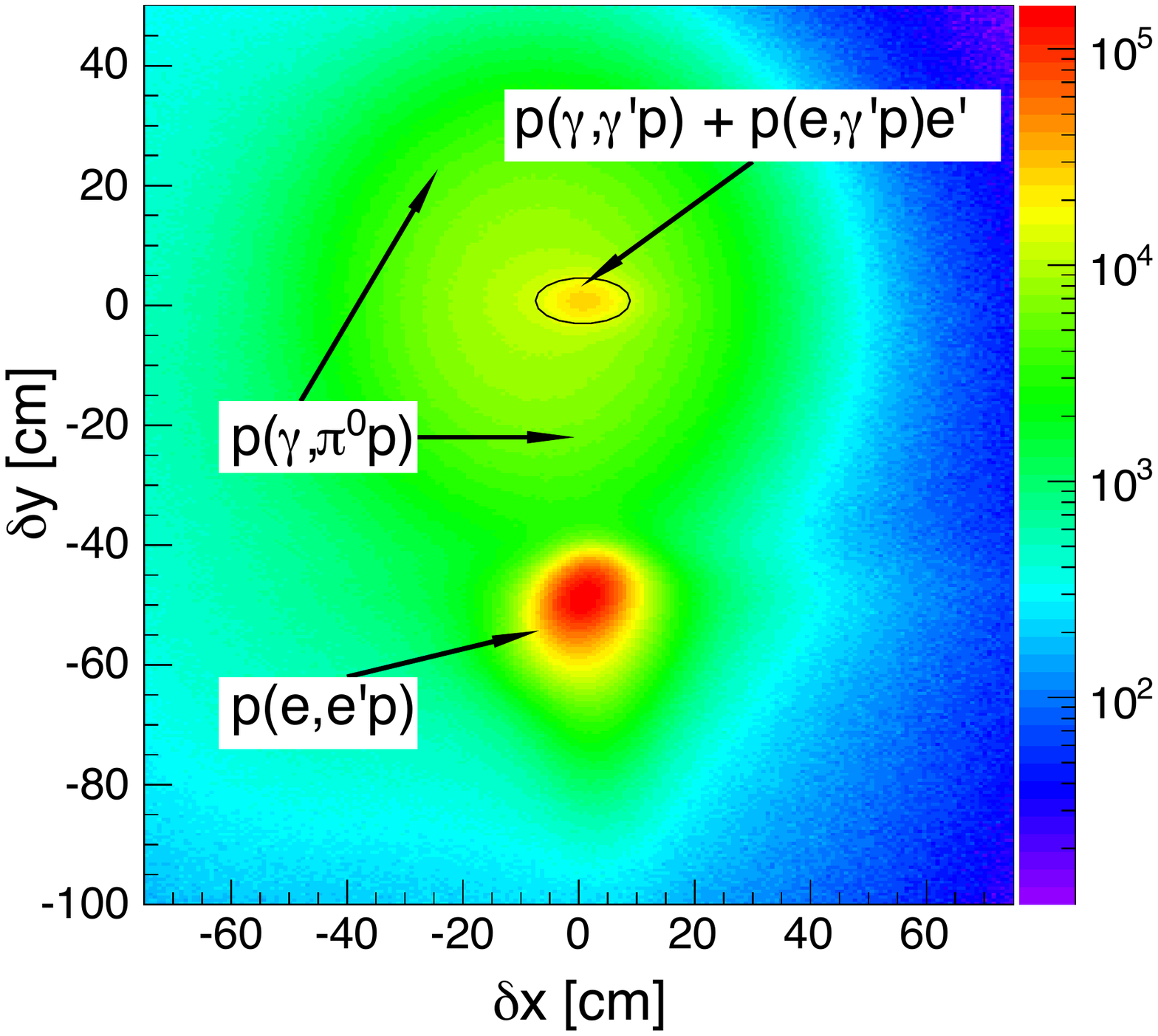}
\vskip -0.1 in
\caption{Two-dimensional distribution of events in $(\delta x: \delta y)$.  
  The WACS events $p(\gamma,\gamma' p)$ and the irreducible Bethe-Heitler
  background $p(e,\gamma'p)e'$ form the peak at $\delta x, \delta y\sim$~0~cm, 
  which is selected by an elliptical cut  \mbox{$(\delta x/8)^2+(\delta y/4)^2 <1$}.
  This region also contains the photopion events
  $p(\gamma,\pi^{_0}p)$ -- the underlying continuum. 
  The $ep$ elastic events are centered at $\delta x\sim$ 0, $\delta y\sim$-50~cm.}
\vskip -0.15 in
\label{fig:dxdy}
\end{figure}

\begin{figure}[!htb]
\vskip -0.0 in
\includegraphics[trim = 100mm 20mm 0mm 25mm, width = 0.52 \textwidth]{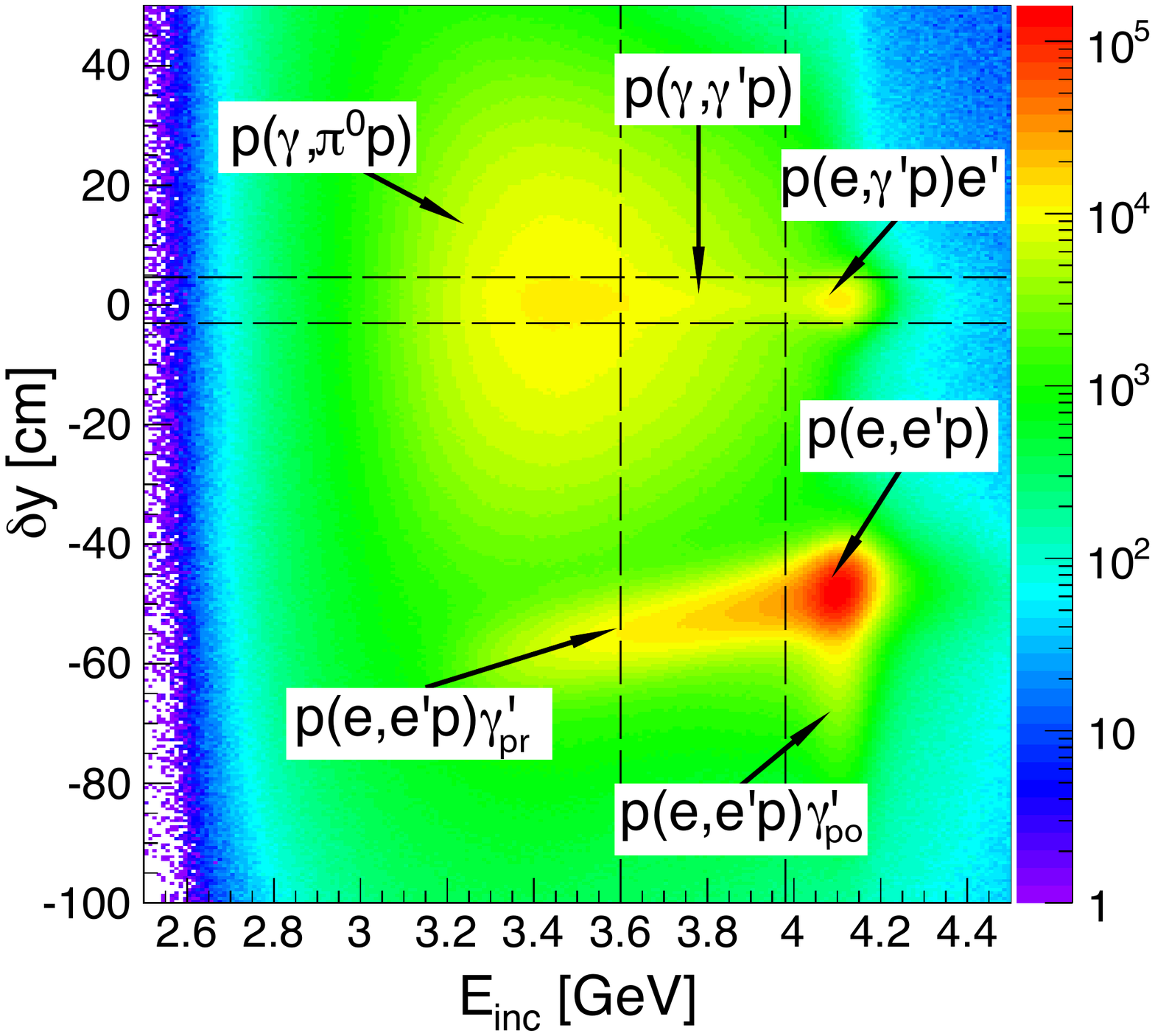}\\
\caption{Two-dimensional distribution of events in (\ein$: \delta y)$.  WACS
  $p(\gamma,\gamma' p)$ events are concentrated around $\delta y\sim$
  0~cm, selected through an elliptical cut in $(\delta x: \delta y)$,
  and reconstructed \ein$\in[3.60,3.98]$~GeV.  The $p(e,\gamma' p)e'$
  events, with a high-energy post scattering radiative photon, are at
  \ein$=\, $4.1~GeV.  Photopion $p(\gamma,\pi^{_{0}}p)$ events are
  mainly located at $\delta y\sim$ 0~cm and \ein$\, < \, $3.6~GeV.  At
  $\delta y$= -50~cm and \ein$\,=\,$4.1~GeV, elastic $p(e,e'p)$ events
  form a peak, with a vertical tail in $\delta y$ of
  $p(e,e'p)\gamma_{po}'$ with a post scattering radiative photon, and
  an oblique tail of $p(e,e'p)\gamma_{pr}'$ with a pre scattering
  radiative photon.}
\vskip -0.15 in
\label{fig:eincdy}
\end{figure}

Potential WACS events are selected based on the kinematic correlation
between the scattered photon and the recoil proton.  The known optical
properties of the HMS are used to reconstruct the momentum, direction,
and reaction vertex of the recoil proton, from which the reconstructed
incident photon (electron) energy (assuming a $\gamma p$ and $ep$
final state), \ein, was determined.  The $\delta x$ and $\delta y$, the
difference in $x$ and $y$ coordinates between the expected and
measured locations of the scattered photon at the entrance of the
calorimeter, were calculated.  The distributions of events in the
$(\delta x: \delta y)$ and (\ein$: \delta y)$ planes are shown in
Figs.~\ref{fig:dxdy} and~\ref{fig:eincdy}, respectively.

The WACS events $p(\gamma,\gamma'p)$, which are concentrated in the
peak at $\delta x, \delta y\sim 0$~cm, lie on top of a continuum
background mainly related to the $p(\gamma,\pi^0 p)$ reaction, for
which one of the photons is detected from the subsequent decay
$\pi^0\rightarrow\gamma\gamma$.  An additional background is due to
electrons and radiative photons from elastic $ep$ scattering.

The recoil proton polarization was measured by the focal plane
polarimeter (FPP) located in the HMS.  The FPP determines the two
polarization components normal to the momentum of the proton by
measuring the azimuthal asymmetries in the angular distribution after
secondary scattering of the proton from an analyzer for positive and
negative electron beam-helicity states.  Two 60~cm (53~g/cm$^{2}$)
thick blocks of CH$_2$ analyzers were used in the experiment.  Two
drift chambers at the focal plane, and a pair of large-acceptance
drift chambers placed after each analyzer, tracked the protons before,
between, and after the analyzer blocks, effectively producing two
independent polarimeters with a combined figure of merit (a product of
efficiency and analyzing power in square) of $7\times10^{-3}$.

\begin{figure}[h]
\centering
\includegraphics[trim = 100mm 50mm 8mm 25mm, width = 0.53 \textwidth]{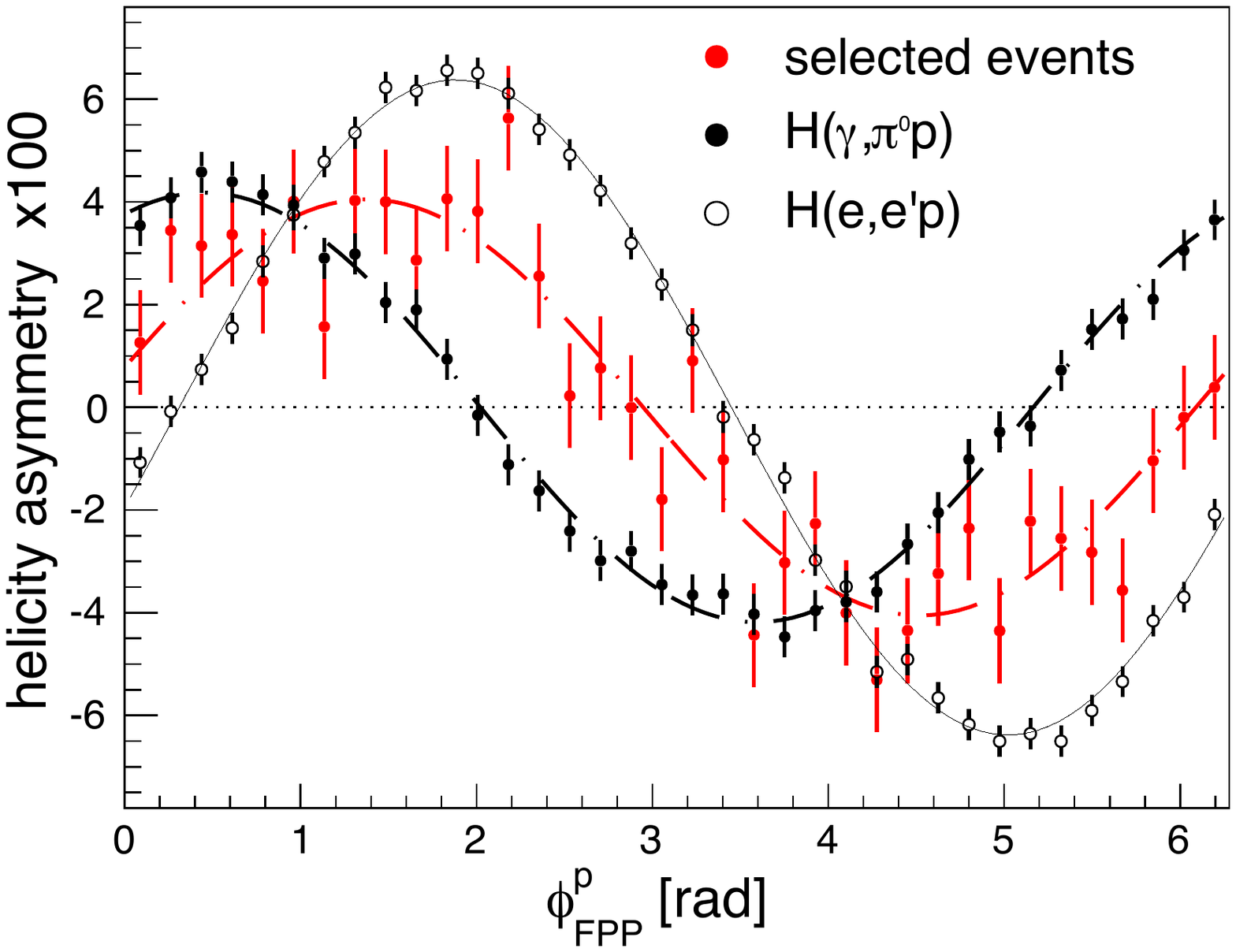}
\vskip -0.25 in
\caption{Azimuthal variation of the difference between beam-helicity
  correlated proton rates.  The selected events (filled red circles)
  are identified with an elliptical cut; see Fig.~\ref{fig:dxdy}.
  Also shown are the corresponding asymmetries for the main background
  events: the elastic $ep$ (open white circles) and photopion
  (filled black circles).}
\label{fig:asymdiff}
\vskip -0.15 in
\end{figure}

For each analyzer, the angular distribution of the scattered protons
is given by
\begin{center}
$N \left( \vartheta, \varphi \right) = N_{0} \left( \vartheta
\right) \left\{ 1 + \left[ A_\mathrm{y} \left( \vartheta \right)
P_{\mathrm{t}}^{\textsc{FPP}} + \alpha \right] \sin \varphi \right. $ \\
$ \left. - \left[ A_\mathrm{y} \left( \vartheta \right) P_\mathrm{n}^{\textsc{FPP}}
+ \beta \right] \cos \varphi \right\} $, \\
\end{center}
where $N_{0}$ is the number of protons that scatter in the
polarimeter, $\vartheta$ and $\varphi$ are the polar and azimuthal
scattering angles, $P_\mathrm{n}^{\textsc{FPP}}$ or
$P_\mathrm{t}^{\textsc{FPP}}$ is the helicity-dependent component of
the proton polarization at the FPP, $A_\mathrm{y}$ is the FPP
analyzing power, and $\alpha$ and $\beta$ are helicity-independent
terms including instrumental asymmetries.  Such a distribution was
measured for the two states of the electron beam helicity.  The
difference between these two distributions, ${N}^{+}/{N_{0} }^{+}
-{N}^{-}/{N_{0}}^{-}$, cancels the instrumental asymmetries to first
order and gives access to the helicity-dependent transferred
polarization.
Performing a Fourier analysis of the beam-helicity difference of $N (
\vartheta, \varphi )$ allows extraction of the products of the proton
polarization components and $A_\mathrm{y}$, shown in
Fig.~\ref{fig:asymdiff}.

Determination of $A_{\mathrm{y}}(\vartheta)$ for each of the analyzers
was performed by measuring the longitudinal polarization of the recoil
proton from $\vec{e}p$ elastic scattering at approximately the same
proton momentum.  This analysis also yields the ratio of the proton
elastic form factors, which was found to be
{\color{black}$\mu_{\mathrm{p}}$\gep/\gmp = 0.744 $\pm$ 0.031} at
\qsq~= 2.25~\gevsq, in excellent agreement with a fit to known
measurements~\cite{pu12}.

Polarization components at the FPP were related to their counterparts
at the target by calculating the proton spin precession in the HMS.
This was done by using a COSY model~\cite{cosy99} of the HMS optics to
obtain a spin transport matrix for each proton track.  The elements of
this matrix are characterized by the average spin precession angle,
which is approximately $100^\circ$.  The proton spin vector was then
transformed to the proton rest frame, with the longitudinal axis
pointing in the direction of the recoil proton in the center-of-mass
frame~\cite{di99}.  In that frame, the longitudinal and sideways
components of the proton polarization normalized to the photon
polarization are just the spin transfer parameters \KLL~and \KLS,
respectively.

The WACS events are selected from a small elliptical region at the
origin of the ($\delta x:\delta y$) plane, as shown in
Fig.~\ref{fig:dxdy}.  For each spin component, the following
deconvolution procedure has been used to extract the final WACS recoil
polarization $K_{_{\textsc{WACS}}}$:
\begin{center}
$K_{_{\mathrm{peak}}} \,\,=\, \left[ K_{_{\mathrm{ellipse}}}-(1-f_{_{1}})K_{_{\mathrm{pion}}}  \right]/f_{_{1}}, $\\
$K_{_{\textsc{WACS}}} \,=\, \left[ K_{_{\mathrm{peak}}} \ \ -(1-f_{_{2}})K_{_{\mathrm{ep}\gamma}} \right] /f_{_{2}},  $\\
\label{eq:prcs}
\end{center}
where $K_{_{\mathrm{ellipse}}}$, $K_{_{\mathrm{peak}}}$,
$K_{_{\mathrm{pion}}}$ and $K_{_{ep\gamma}}$ are the polarizations
related to (i) all of the events within the ($\delta x:\delta y$)
ellipse shown in Fig.~\ref{fig:dxdy}, (ii) only the events in the peak
above the continuum background, (iii) the pion photoproduction
background events, and (iv) the Bethe-Heitler background events.
The fractions for the ratio of event types are defined as
$f_{_{1}}\,=\,N_{_{\mathrm{peak}}}/N_{_{\mathrm{ellipse}}}$ and
$f_{_{2}}=N_{_{\textsc{WACS}}}/N_{_{\mathrm{peak}}}$, respectively.
The dominant background polarization, by itself an important physics
result, $K_{_{\mathrm{pion}}}$, was measured by selecting events from
regions of the ($\delta x:\delta y$) plane in Fig.~\ref{fig:dxdy} that
contain neither WACS nor $ep$ ($ep\gamma$) events, corresponding to
$\delta y > \, $10~cm and \mbox{-35$\, < \delta y < \, $-10~cm.}  It
was found that within the statistical precision of the measurements,
$K_{_{\mathrm{pion}}}$ was constant over broad regions of that plane.
The polarization of the $ep\gamma$ background $K_{_{\mathrm{ep}\gamma}}$ 
was determined by selecting events in the deflected $ep$ elastic peak region. 
It was determined that within the statistical precision, the polarization 
$K_{_{\mathrm{ep}}}$ was consistent with $K_{_{\mathrm{ep}\gamma}}$. 
The results for the longitudinal and sideways components are
\mbox{$K_{_{\mathrm{ep}\gamma, \textsc{LL}}}$ = 0.4513 $\pm$ 0.0054}
and \mbox{$K_{_{\mathrm{ep}\gamma, \textsc{LS}}}$ = -0.1837 $\pm$
  0.0054.}  Systematic uncertainties arising from the methods used to
determine the background polarization observables have been studied
and included in the final results.  The final source of background
which needs to be taken into account arises as a result of the
presence of accidental random events in the final sample ($<$4\% of
the events).

Results obtained with the two polarimeters were statistically
consistent and were combined to form a weighted mean.  With the WACS
region selected to obtain the best statistical accuracy on
$K_{_{\textsc{WACS}}}$ and fits performed on the respective
distributions, we find \mbox{$f_{_{1}}=0.405 \pm 0.004$.}  The
determination of the fraction $f_{_{2}}$ is a little more involved and
requires analysis of calibration data taken without the copper
radiator.  By doing so, we measure the quantity $n_{\mathrm{ep}\gamma}
= N_{_{\mathrm{ep}\gamma}}/{N_{_{\mathrm{ep}}}}$ after having imposed
an optimized cut on the incident energy to remove prevertex
interactions, which can then be used to determine the fraction
\mbox{$f_{_{2}}=1-
  {N_{\mathrm{ep}\gamma}}/{N_{_{peak}}}=1-n_{\mathrm{ep}\gamma} \times
  {N_{_{\mathrm{ep}}}}/{N_{_{\mathrm{peak}}}}$} for the production
data.  Following this method, we find \mbox{$f_{_{2}}=$0.67 $\pm$
  0.03.}  As a consistency check, this fraction was determined using a
second analysis method which involved reconstructing the incident
energy spectrum for $ep\gamma$ events through a convolution of the
theoretical bremsstrahlung spectrum from the radiator with the real
spectrum of calibration events (without the radiator) for a scattered
photon detected at $\delta y \sim$ 0~cm.  A third method consists of
extracting the fraction from the difference between the value of the
cross section measured in this experiment, which includes the
irreducible $ep\gamma$ background, and the value of the WACS cross
section obtained from the parametrization of the E99-114
results~\cite{da07}.  The results obtained for $f_{_{2}}$ using these
three methods were found to be consistent with each other.

\begin{table}[h]
\vskip -0.08 in
  \caption{The WACS and pion photoproduction polarization results. In this experiment
  the values of the Mandelstam variables are \invs = 7.8 (0.3) and \invt = -2.1 (0.1) \gevsq, 
  which define \cma = 70 (2)$^\circ$. The values in parentheses are the acceptance ranges. 
  For the WACS polarization the first uncertainty is statistical and the second is systematic.
  For the pion photoproduction polarization the combined uncertainty is shown.
  In E99-114 (Ref.~\cite{ha05}), $s$ = 6.9 \gevsq~and  \cma = 120$^{\circ}$.}
\begin{ruledtabular}
\renewcommand{\arraystretch}{1.2}
\begin{tabular}{l c c}
Selection	& $K_{_{\mathrm{LL}}}$		& $K_{_{\mathrm{LS}}}$	\\  
\hline
Ellipse	& 0.180$\pm$0.015				& $-$0.233$\pm$0.015 \\
Peak		& 0.565$\pm$0.038				& $-$0.142$\pm$0.038 \\
WACS$_{\mathrm{_{this \,\, experiment}}}$	& 0.645$\pm$0.059$\pm$0.048	& $-$0.089$\pm$0.059$\pm$0.040 \\
WACS$_{\mathrm{_{E99-114}}}$	& 0.678$\pm$0.083$\pm$0.04	& 0.114$\pm$0.078$\pm$0.04 \\
Pion$_{\mathrm{_{this \,\, experiment}}}$	& $-$0.082$\pm$0.007	& $-$0.296$\pm$0.007 \\
Pion$_{\mathrm{_{E99-114}}}$	& 0.532$\pm$0.006		& 0.480$\pm$0.006 \\
\end{tabular}
\end{ruledtabular}
\label{tab:results}
\vskip -0.15 in
\end{table}

The extracted polarization transfer observables for different event
samples are given in Table~\ref{tab:results}.  A comparison between
the E99-114~\cite{ha05} and the present polarization measurement
results for WACS and a single-pion photoproduction is also given in
Table~\ref{tab:results}.  The large changes of \KLL, \KLS~in pion
production between the two data sets indicate a complicated
nonasymptotic reaction mechanism.  These results are in good
agreement with previous measurements~\cite{wij02}.

\begin{figure}[!htb]
\vskip -0.10 in
\includegraphics[trim = 5mm 0mm 0mm 0mm, width = 0.49\textwidth]{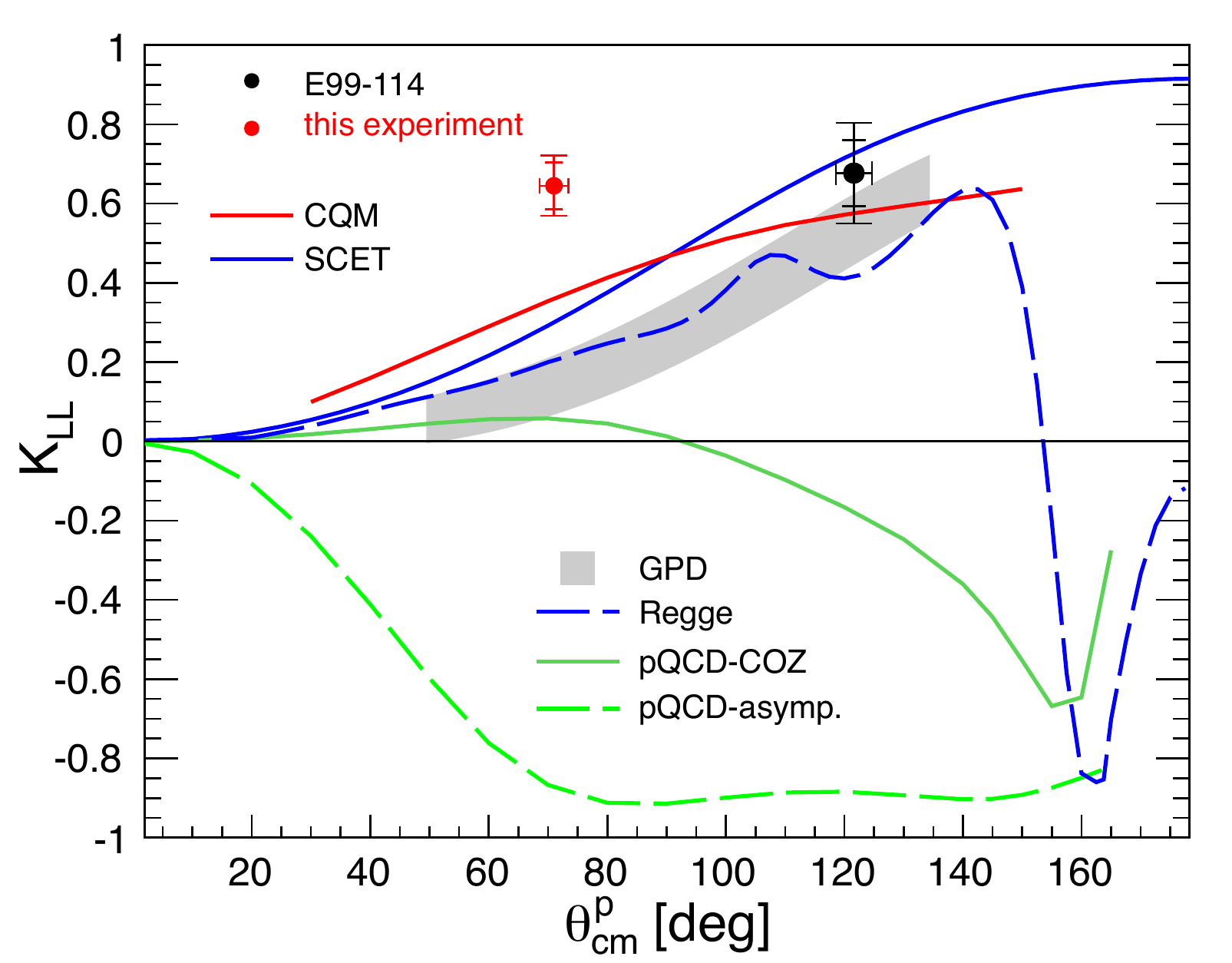}
\caption{The experimental result for \KLL. 
Also shown are the E99-114 value~\cite{ha05} and 
the calculations in different approaches: the pQCD~\cite{kr91} with the asymptotic and COZ 
distribution amplitudes~\cite{ch89}, the extended Regge model~\cite{ca03}, 
the GPD~\cite{hu02}, shown as a gray band of uncertainty due to finite mass 
corrections~\cite{di03}, the CQM~\cite{mi04}, and the SCET~\cite{ki13}.}
\label{fig:kll}
\vskip -0.10 in
\end{figure}

The final result for WACS \KLL~is shown in Fig.~\ref{fig:kll} along
with the predictions of the relevant calculations.  In agreement with
what was found in the previous JLab experiment~\cite{ha05}, the
\KLL~result is inconsistent with predictions based on the pQCD
theory~\cite{kr91} and hence suggests that even at this experiment's
values of \mbox{$s, -t, -u \,=\, 7.8, \, 2.1, \, 4.0$~\gevsq} we are
still far from the asymptotic regime for the WACS process.
Figure~\ref{fig:kll} shows the pQCD predictions for the two extreme
choices of the distribution amplitudes, namely the asymptotic and
COZ~\cite{ch89}, with the first, \ada, for asymptotically large energy
scales, and the second having a peak in $\phi_1$ at $x_{1,3} \approx
1/2$, $x_2 \approx 0$, to be constrained by the QCD-sum-rules-based
values of its lowest moments.

In conclusion, the polarization transfer observables \KLL~and
\KLS~have been measured for proton Compton scattering at a new
kinematic point at \mbox{\invs =7.8~\gevsq} and \cma$ = 70^\circ$.
The final results are \mbox{\KLL$= 0.645 \pm 0.059 \pm 0.048$} and
\KLS$ = -0.089 \pm 0.059 \pm 0.040$, where the first uncertainty is
statistical and the second is systematic.  The \KLS~result is in
agreement within the experimental uncertainties with calculations for
both the leading-quark and the pQCD approaches~\cite{hu02, mi04, ki13,
  kr91} and hence suggests that there is no strong evidence for proton
helicity flip in this reaction.  The value obtained for \KLL~is, quite
unexpectedly, larger than all the available theoretical predictions.
Such a \KLL~could be caused by noncollinear effects in exclusive
reactions at currently accessible energies and parton correlations in
the nucleon.  In this respect, the \KLL~increase may be related to
significant roles observed in elastic electron-nucleon scattering of
both quark orbital angular momentum and a $u-d$ diquark
correlation~\cite{jones00, CJRW}.

We thank the Jefferson Lab Hall C technical staff for their
outstanding support.  This work was supported in part by the INFN
gruppo Sanit\`{a}, the National Science Foundation, the UK Science and
Technology Facilities Council (STFC 57071/1, 50727/1), and the DOE
under Contract No.~DE-AC05-84ER40150 Modification No. M175, under which
the Southeastern Universities Research Association (SURA) operates the
Thomas Jefferson National Accelerator Facility.


\begin{references}
%
\bibitem{GPDintro} D.~M$\ddot{\textup{u}}$ller, D.~Robaschik, B.~Geyer, 
F.-M.~Dittes, and J.~Ho\u{r}ej\u{s}i, {\em Fortschr. Phys.} {\bf 42}, 101 (1994); 
X.D.~Ji, \PRL {\bf 78}, 610 (1997); A.V.~Radyushkin, {\it Phys. Lett.} B {\bf 380}, 417 (1996).
%
\bibitem{GPD_hadron} A.V.~Belitsky and A.V.~Radyushkin, {\em Phys. Rep.} {\bf 418}, 1 (2005).
%
\bibitem {fa90} G.R.~Farrar and H.~Zhang, \PRL {\bf 65}, 1721 (1990); {\em Phys.~Rev.~D} {\bf 41}, 3348 (1990).
%
\bibitem {kr91}A.S.~Kronfeld and B.~Nizic, {\em Phys.~Rev.~D} {\bf 44}, 3445 (1991); 
M.~Vanderhaeghen, P.A.M.~Guichon, and J.~Van~de~Wiele, \NP {\bf A622}, 144c (1997);
T.~Brooks and L.~Dixon, {\em Phys.~Rev.~D} {\bf 62}, 114021 (2000).
%
\bibitem{ra98} A.V.~Radyushkin, {\em Phys.~Rev.~D} {\bf 58}, 114008 (1998).
%
\bibitem{di99} M.~Diehl, T.~Feldmann, R.~Jakob, and P.~Kroll, {\em Eur.~Phys.~J.~C} {\bf 8}, 409 (1999).
%
\bibitem{hu02} H.W.~Huang, P.~Kroll, and T.~Morii, {\em Eur.~Phys.~J.~C} {\bf 23}, 301 (2002); 
{\bf 31}, 279(E) (2003).
%
\bibitem{mi04} G.A.~Miller, {\em Phys.~Rev.~C} {\bf 69}, 052201(R) (2004).
%
\bibitem{ca03} F.~Cano and J.M.~Laget, {\em Phys. Lett.~B} {\bf 551}, 317 (2003).
%
\bibitem{ki13} N.~Kivel and M.~Vanderhaeghen, \JHEP 04 (2013) 29;
\NP {\bf B883}, 224 (2014); arXiv:1504.00991.
%
\bibitem{br73} S.J.~Brodsky and G.~Farrar, \PRL {\bf 31}, 1153 (1973);
V.A.~Matveev, R.M.~Muradyan, and A.V.~Tavkheldize, {\em Lett. Nuovo Cimento} { \bf7}, 719 (1973).
%
\bibitem{sh79} M.A.~Shupe \etal, {\em Phys.~Rev.~D}  {\bf 19}, 1921 (1979).
%
\bibitem{da07} A.~Danagoulian \etal, \PRL {\bf 98}, 152001 (2007).
%
\bibitem{ha05} D.J.~Hamilton \etal, \PRL {\bf 94}, 242001 (2005).
%
\bibitem{pu12} 	A.J.R.~Puckett \etal, {\em Phys.~Rev.~C} {\bf 85}, 045203 (2012);
  V.~Punjabi \etal, {\em Phys.~Rev.~C} {\bf 71}, 055202 (2005); {\em Phys.~Rev.~C} {\bf 71}, 069902 (2005).
%
\bibitem{cosy99} K.~Makino and M.~Berz, \NIM Sect. A {\bf 427}, 338 (1999).
%
\bibitem{wij02} K.~Wijesooriya \etal,  {\em Phys.~Rev.~C}  {\bf 66}, 034614 (2002); 
W.~Luo \etal, \PRL {\bf 108}, 222004 (2012).
%
\bibitem{di03} M.~Diehl, Th.~Feldmann, H.W.~Huang, and P.~Kroll, \PR D {\bf 67}, 037502 (2003).
%
\bibitem{ch89} V.L.~Chernyak, A.A.~Oglobin, and A.R.~Zhitnitsky, \mbox{ {\em Z. Phys. C} {\bf 42}, 569 (1989).}
%
\bibitem{jones00} M.~Jones \etal, \PRL {\bf 84}, 1398 (2000);
O.~Gayou \etal, \PRL {\bf 88}, 092301 (2002);
G.A.~Miller, \PR C {\bf 66}, 032201(R) (2002).
%
\bibitem{CJRW} G.D.~Cates, C.~W.~de Jager, S.~Riordan, and B.~Wojtsekhowski, \PRL {\bf 106}, 252003 (2011);
C.D.~Roberts, M.S.~Bhagwat, A.~H\"{o}ll, and S.V.~Wright, Eur. Phys. J. Spec. Top. {\bf 140}, 53 (2007); 
I.C.~Clo\"et, G.~Eichmann, B.~El-Bennich, T.~Kl\"{a}hn, and C.D.~Roberts, Few-Body Syst. {\bf 46}, 1 (2009).
%
\end{references}
\end{document}